\documentclass[reprint,twocolumn,superscriptaddress,showpacs,showkeys,preprintnum bers,amsmath,amssymb,floatfix,aps,pra,longbibliography]{revtex4-1}

\usepackage{graphicx}
\usepackage{bm}

\usepackage{placeins}
\usepackage{graphics}
\usepackage{color}
\usepackage{hyperref}
\usepackage{multirow}
\usepackage{blindtext}
\usepackage{appendix}
\usepackage{xcolor}
\hypersetup{
    colorlinks,
    linkcolor={red!50!black},
    citecolor={blue!50!black},
    urlcolor={blue!80!black}
}

\begin{document}


\title{Electronic and Spin-Orbit Properties of hBN Encapsulated Bilayer Graphene}

\author{Klaus Zollner}
	\email{klaus.zollner@physik.uni-regensburg.de}
	\affiliation{Institute for Theoretical Physics, University of Regensburg, 93053 Regensburg, Germany}	
  \author{Eike Icking}
\affiliation{JARA-FIT and 2nd Institute of Physics, RWTH Aachen University, 52074 Aachen, Germany,~EU}%
\affiliation{Peter Gr\"unberg Institute  (PGI-9), Forschungszentrum J\"ulich, 52425 J\"ulich,~Germany,~EU}
\author{Jaroslav Fabian}
	\affiliation{Institute for Theoretical Physics, University of Regensburg, 93053 Regensburg, Germany}

\date{\today}

\begin{abstract}
Van der Waals (vdW) heterostructures consisting of Bernal bilayer graphene (BLG) and hexagonal boron nitride (hBN) are investigated. 
By performing first-principles calculations we capture the essential BLG band structure features for several stacking and encapsulation scenarios. A low-energy model Hamiltonian, comprising orbital and spin-orbit coupling (SOC) terms, is employed to reproduce the hBN-modified BLG dispersion, spin splittings, and spin expectation values. Most important, the hBN layers open an orbital gap in the BLG spectrum, which can range from zero to tens of meV, depending on the precise stacking arrangement of the individual atoms. Therefore, large local band gap variations may arise in experimentally relevant moir\'{e} structures.
Moreover, the SOC parameters are small (few to tens of $\mu$eV), just as in bare BLG, but are markedly proximity modified by the hBN layers. 
Especially when BLG is encapsulated by monolayers of hBN, such that inversion symmetry is restored, the orbital gap and spin splittings of the bands vanish.
In addition, we show that a transverse electric field mainly modifies the potential difference between the graphene layers, which perfectly correlates with the orbital gap for fields up to about 1~V/nm. Moreover, the layer-resolved Rashba couplings are tunable by $\sim 5~\mu$eV per V/nm.
Finally, by investigating twisted BLG/hBN structures, with twist angles between 6$^{\circ}$ -- 20$^{\circ}$, we find that the global band gap increases linearly with the twist angle. The extrapolated $0^{\circ}$ band gap is about 23~meV and results roughly from the average of the stacking-dependent local band gaps.
Our investigations give new insights into proximity spin physics of hBN/BLG heterostructures, which should be useful for interpreting experiments on extended as well as confined (quantum dot) systems. 
\end{abstract}

\maketitle

\section{Introduction}

Two-dimensional (2D) materials provide a fascinating playground for investigating fundamental physics phenomena and serve as promising platforms for technological applications~\cite{Cai2018:CSR}, such as 
tunnel field-effect transistors based on monolayer transition-metal dichalcogenides (TMDCs)~\cite{Kumar2020:S} or magnetic random access memories
based on 2D magnets~\cite{Zhang2019:IM,Gong2019:SC,Li2019:AM,Song2018:SC}.
When different 2D materials are combined in van der Waals (vdW) heterostructures, new device functionalities can be engineered by the proximity effect~\cite{Sierra2021:NN}.

Bernal bilayer graphene (BLG) is a prototypical electronic material which offers high electron mobility, like monolayer graphene, but also band gap engineering by an external electric displacement field~\cite{McCann_2006, Min_2007,McCann2013:RPP, Slizovskiy_2021}. In experiments, the band gap can be tuned up to hundreds of meV~\cite{Ohta_2006,Zhou_2007,Oostinga_2007,Mak_2009,Icking2022:AEM}, while the bands additionally host magnetic and superconducting phases~\cite{Zhou_2022,Barrera_2022}. The tunable flat bands make BLG also an interesting platform for studying electron-electron interactions~\cite{Weitz_2010,geisenhof_2021,Kou_2014,Ki_2014,Zheng_2020,Seiler_2022}.
Furthermore, BLG has low spin-orbit coupling (SOC)~\cite{Kane_2005,Konschuh2012:PRB,Kurzmann_2021,Banszerus_2021} and weak hyperfine interaction~\cite{Wojtaszek_2014, Fischer_2009}, which, in combination with the possibility of electrostatic confinement of charge carriers in quantum point contacts and quantum dots~\cite{Overweg_2018,Kurzmann_2021,Eich_2018,Banszerus_2020,Garreis_2021,Lee_2020}, makes it an attractive candidate for hosting spin and valley qubits~\cite{Banszerus_2023}.

Furthermore, BLG is highly sensitive to its environment. For example, BLG on a strong spin-orbit material, such as TMDCs or topological insulators (TIs), exhibits enhanced spin interactions in the meV range~\cite{Zutic2019:MT,Zollner2018:NJP, Zollner2019:PRB,Zollner2020:PRL, Zollner2019b:PRB, Omar_2019, Bisswanger_2022,Gmitra2017:PRL}. In particular, for BLG on WSe$_2$ it has been theoretically predicted~\cite{Gmitra2017:PRL} and experimentally confirmed~\cite{Island2019:Nat}, that the SOC of charge carriers can be controlled by the displacement field leading to spin-orbit valve functionality.
It has also been shown that WSe$_2$ stabilizes the superconducting phase in BLG due to proximity-induced Ising SOC~\cite{zhang_2023}. In addition to the SOC, it is also possible to induce superconductivity~\cite{Li_2020,Moriya2020:PRB} or magnetism~\cite{Zollner2016:PRB, Zollner2018:NJP,Zollner2019:PRB} in BLG with high tunability~\cite{Zollner2021:PRB}, making it an intriguing material for various applications.

Encapsulating BLG in hexagonal boron nitride (hBN) has become a standard technique for improving its quality and stability~\cite{Dean_2010,Wang_2013}, due to the insulating nature of hBN with a band gap of 5.9~eV~\cite{Watanabe_2004}, its high thermal conductivity~\cite{Lindsay_2011,Yuan_2019}, and its ultra-flat surface. This technique has made it possible to build ultra-high carrier mobility devices based on graphene~\cite{Dean_2010,Wang_2013,Banszerus2015:SA}, as well as high-quality optoelectric TMDC-devices~\cite{Ajayi_2017,Raja_2019,Ersfeld_2020}, while also protecting sensitive materials from degradation under ambient conditions~\cite{Gao_2016,Gurram_2018}. Improving the quality of hBN is a focus of ongoing research efforts~\cite{Meng_2019,Maestre_2021,Taniguchi_2007,Zastrow_2019,onodera_2019,Onodera_2020,Kubota_2007,Kubota_2008,hoffmann_2014,edgar_2014,Liu_2018,Li_2020hBN,Li_2021,Zhang_2019,Cao_2022,Naclerio_2022,Ouaj_2023}.
In addition to its encapsulation properties, recent research has shown that hBN can also have a profound effect on the bandstructure of BLG, breaking the inversion symmetry and allowing for manipulation of the induced Berry curvature~\cite{Kareekunnan2020:PRB,Moulsdale2022:PRB}. 

In this paper we investigate theoretically, using DFT methods, the orbital and spin-orbit proximity effects in BLG/hBN and fully encapsulated hBN/BLG/hBN heterostructures. 
The calculated electronic structures are fitted---at low energies---with effective Hamiltonians and various quantitative parameters of the proximitized bands are extracted. We analyze the variation of the 
proximity band structure with stacking/encapsulation configurations, finding 
orbital gaps up to tens of meV. The spin-orbit splittings of bands are small, tens of $\mu$eV, consistent with expectations, and with previous calculations of hBN-encapsulated monolayer graphene \cite{Zollner2019:PRB}. We pay particular attention to the electric field tunability. While the electric displacement field opens a large gap, on the order of
50--100 meV for fields of 1V/nm, the field also polarizes the layers which then leads to layer-selective proximity effect. For example, the Rashba SOC is strongly influenced by the electric field, changing the sign as the field changes the polarity. We also find a rather strong enhancement of the interlayer spin-orbit coupling parameter, from 10 to more than 20 $\mu$eV, due to encapsulation. Finally, we also study orbital gaps in twisted BLG/hBN heterostructures, to see spatial variation of the gaps, resolving the
individual atomic orbital contributions in the spatial profile. 

Although the SOC effects due to hBN are rather modest, experimentally it is now possible 
to resolve $\mu$eV scales, most prominently in confined quantum dot structures. The experiments find SOC gaps of about 40-80~$\mu$eV in hBN encapsulated BLG \cite{Banszerus_2020}, while the theoretical spin-orbit splittings of the infinite crystals are up to 25 $\mu$eV~\cite{Konschuh2012:PRB}. While this discrepancy is not fully resolved, our calculations should be useful to interpret also such experiments on confined structures. 

The paper is organized as follows: In section~\ref{sec:comp} we address the structural setup and summarize the calculation details for obtaining the electronic structures. In section~\ref{sec:model}, we introduce the model Hamiltonians that capture the low-energy physics of proximitized BLG (including orbital and SOC terms), which are used to fit the DFT-calculated dispersions. In section~\ref{sec:results}, we then show and discuss the DFT-calculated electronic structures of (hBN)/BLG/hBN stacks, along with the model Hamiltonian fits, as well as the gate tunability.
In section~\ref{sec:twist}, we study twisted BLG/hBN structures and discuss local band gap variations in moir\'{e} geometries.
Finally, in section~\ref{sec:summ} we conclude the manuscript.

\section{Computational Details and Geometry}
\label{sec:comp}
\begin{figure}[!htb]
	\includegraphics[width=0.8\columnwidth]{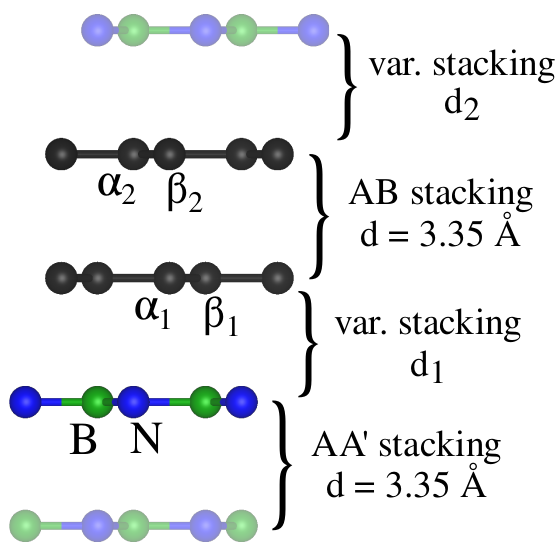}
	\caption{Side view of BLG encapsulated by hBN layers. The interlayer distances and stackings within BLG and hBN are fixed. Stackings between the graphene and hBN layers are variable with interlayer distances d$_1$ and d$_2$.
 \label{Fig:Structure}}
\end{figure}

In order to calculate the electronic band structure of (hBN)/BLG/hBN heterostructures, we use a common unit cell for graphene and hBN. 
Therefore, we fix the lattice constant of graphene \cite{Neto2009:RMP} to $a = 2.46$~\AA, and change the hBN lattice constant, from its experimental value \cite{Catellani1987:PRB} 
of $a = 2.504$~\AA, to the graphene one. The lattice constants of graphene and hBN differ by less than 2\%, justifying our theoretical considerations of commensurate geometries.
Experimentally, the small lattice mismatch does lead to moir\'{e} patterns \cite{Jung2014:PRB,Moon2014:PRB,Argentero2017:NL}.
Here we consider several, but not all, structural arrangements for commensurate unit cells, so as to 
get a quantitative feeling for spin-orbit phenomena in a generic experimental setting. 

In the following, we consider BLG in Bernal (AB) stacking with a fixed interlayer distance of 3.35~\AA~\cite{Konschuh2012:PRB}. The minimal heterostructure we study is BLG on top of monolayer hBN in different stacking scenarios, see Fig.~\ref{Fig:Structure}. In addition, we consider BLG encapsulated by hBN layers. When bilayer hBN is considered for encapsulation, we employ an AA' stacking also with a fixed interlayer distance of 3.35~\AA~\cite{Constantinescu2013:PRL}. 
The variable stackings are between the graphene layers and the surrounding hBN layers, see Fig.~\ref{Fig:Structure}. We use the equilibrium distances of the individual graphene/hBN stackings as found in Ref.~\cite{Zollner2019:PRB}. This fixes all interlayer distances in our (hBN)/BLG/hBN heterostructure geometries for all stacking sequences we consider.

Before we proceed, we define a terminology to make sense of the structural arrangements used in the following. We denote the three relevant sites in hBN as the B-site (Boron), the N-site (Nitrogen), and the H-site (hollow position in the center of the hexagon). 
Similarly, each graphene layer, $j = \{1,2\}$, consists of sublattices $\alpha_j$ (C$_{\textrm{A}}$) and $\beta_j$ (C$_{\textrm{B}}$) and a hollow site h$_j$.
The energetically most favorable stacking sequence, listed from bottom to top, of BLG on monolayer hBN is then abbreviated as (B$\beta_1$h$_2$,H$\alpha_1\beta_2$), see Fig.~\ref{Fig:Structure}. This stacking sequence has d$_1 = 3.35$~\AA~\cite{Zollner2019:PRB}.
In Table~\ref{Tab:Geometries}, we summarize the total energies and interlayer distances of all our investigated geometries. 
In addition, we consider all hBN-encapsulated structures also with bilayer-hBN below BLG (not explicitly listed in Table~\ref{Tab:Geometries}).

\begin{table}[!htb]
\begin{ruledtabular}
\caption{\label{Tab:Geometries} Summary of total energies and interlayer distances for (hBN)/BLG/hBN heterostructures.}
\begin{tabular}{lccc}
Configuration & E$_{\textrm{tot}}$-E$_0$ [meV] & d$_1$ [\AA] & d$_2$ [\AA]\\\hline
(B$\beta_1$h$_2$,H$\alpha_1\beta_2$) & 0 & 3.35 & -\\
(N$\beta_1$h$_2$,H$\alpha_1\beta_2$) & 13.911 & 3.50 & -\\
(N$\beta_1$h$_2$,B$\alpha_1\beta_2$) & 16.709 & 3.55 & - \\\hline
(B$\beta_1$h$_2$N,H$\alpha_1\beta_2$H) & 0 & 3.35 & 3.35 \\
(N$\beta_1$h$_2$B,H$\alpha_1\beta_2$H) & 27.745 & 3.50 & 3.50 \\
(B$\beta_1$h$_2$B,H$\alpha_1\beta_2$H) & 13.791 & 3.35 & 3.50 \\
(B$\beta_1$h$_2$H,H$\alpha_1\beta_2$B) & 16.602 & 3.35 & 3.55 \\
(N$\beta_1$h$_2$H,H$\alpha_1\beta_2$B) & 30.556 & 3.50 & 3.55 \\
(N$\beta_1$h$_2$H,B$\alpha_1\beta_2$B) & 33.361 & 3.55 & 3.55 \\
\end{tabular}
\end{ruledtabular}
\end{table}

Initial atomic structures are set up with the atomic simulation environment (ASE) \cite{ASE} and visualized with VESTA software \cite{VESTA}.
First-principles calculations are performed with 
full potential linearized augmented plane wave (FLAPW) code 
based on DFT \cite{Hohenberg1964:PRB} and implemented in \texttt{WIEN2k} \cite{Wien2k}. 
Exchange-correlation effects are treated with the 
generalized-gradient approximation (GGA) \cite{Perdew1996:PRL}, including dispersion 
correction \cite{Grimme2010:JCP} and using a $k$-point grid of 
$48\times 48 \times 1$ in the hexagonal Brillouin-Zone if not specified otherwise. 
The values of the Muffin-tin radii we use are $r_{\textrm{C}}=1.34$ for C atom, 
$r_{\textrm{B}}=1.27$ for B atom, and $r_{\textrm{N}}=1.40$ for N atom. 
We use the plane wave cutoff parameter $RK_{\textrm{MAX}} = 9.5$.
In order to avoid interactions between periodic images of our slab geometry, we add a vacuum of at least $30$~\AA~in $z$-direction.  

For some of our structures, we also study the effect of a transverse electric field. 
Importantly, in \texttt{WIEN2k} the electric field is modeled by a zigzag potential~\cite{Stahn2001:PRB}, that is realized by a Fourier series, and which has its discontinuities at $z = 0$ and $z = 0.5c$, where $c$ is the lattice constant in $z$-direction. Therefore, in all our calculations we center our geometries around $z = 0.25c$, similar as in Ref.~\cite{Gmitra2009:PRB}. A positive (negative) electric field corresponds to a decreasing (increasing) potential across our geometries.

\section{Model Hamiltonian}
\label{sec:model}
Here we present the Hamiltonian used to model the low-energy bands of the encapsulated BLG structures. The basis states are
$|\textrm{C}_{\textrm{A1}}, \uparrow\rangle$, $|\textrm{C}_{\textrm{A1}}, \downarrow\rangle$, 
$|\textrm{C}_{\textrm{B1}}, \uparrow\rangle$, $|\textrm{C}_{\textrm{B1}}, \downarrow\rangle$, 
$|\textrm{C}_{\textrm{A2}}, \uparrow\rangle$, $|\textrm{C}_{\textrm{A2}}, \downarrow\rangle$, 
$|\textrm{C}_{\textrm{B2}}, \uparrow\rangle$, and $|\textrm{C}_{\textrm{B2}}, \downarrow\rangle$.
In this basis the Hamiltonian is (see Ref. \cite{Konschuh2012:PRB})
\begin{flalign}
\mathcal{H} = & \mathcal{H}_{\textrm{orb}} + \mathcal{H}_{\textrm{soc}}+E_D,\\
\mathcal{H}_{\textrm{orb}} = & \begin{pmatrix}
\Delta+V & \gamma_0 f(\bm{k}) & \gamma_4 f^{*}(\bm{k}) & \gamma_1 \\
\gamma_0 f^{*}(\bm{k}) & V & \gamma_3 f(\bm{k}) & \gamma_4 f^{*}(\bm{k}) \\
 \gamma_4 f(\bm{k}) & \gamma_3 f^{*}(\bm{k}) & -V & \gamma_0 f(\bm{k}) \\
\gamma_1 & \gamma_4 f(\bm{k}) & \gamma_0 f^{*}(\bm{k}) & \Delta-V
\end{pmatrix} \otimes s_0,\\
 \mathcal{H}_{\textrm{soc}}= &
\begin{pmatrix}
\tau \lambda_{\textrm{I}}^\textrm{A1}s_z & 2\textrm{i}\lambda_{\textrm{R1}}s_{-}^{\tau} & \textrm{i}\lambda_4^\textrm{A}s_{+}^{\tau} & 0\\
-2\textrm{i}\lambda_{\textrm{R1}}s_{+}^{\tau} & -\tau \lambda_{\textrm{I}}^\textrm{B1}s_z & 0 & -\textrm{i}\lambda_4^\textrm{B}s_{+}^{\tau}\\
-\textrm{i}\lambda_4^\textrm{A}s_{-}^{\tau} & 0 & \tau \lambda_{\textrm{I}}^\textrm{A2}s_z & 2\textrm{i}\lambda_{\textrm{R2}}s_{-}^{\tau}\\
0 & \textrm{i}\lambda_4^\textrm{B}s_{-}^{\tau} & -2\textrm{i}\lambda_{\textrm{R2}}s_{+}^{\tau} & -\tau \lambda_{\textrm{I}}^\textrm{B2}s_z
\end{pmatrix}.
\end{flalign}
Here, $\gamma_j$, $j = \{ 0,1,3,4 \}$, describe intra- and interlayer hoppings in BLG.
The parameter $\gamma_0$ is the nearest neighbor intralayer hopping, similar to monolayer graphene, while $\gamma_1$ is the direct interlayer hopping. The parameters $\gamma_3$ and $\gamma_4$ describe indirect hoppings between the layers. 
The vertical hopping $\gamma_1$ connects only 2 atoms and therefore it appears without structural function in the Hamiltonian. 
In contrast, the other hoppings couple an atom to 3 corresponding nearest neighbor partner atoms, hence they appear with structural function, where we use the linearized version, 
$f(\bm{k}) = -\frac{\sqrt{3}a}{2}(\tau k_x-\textrm{i}k_y)$, valid in the vicinity of the K points \cite{Kochan2017:PRB}. The graphene lattice constant is
$a$ and the Cartesian wave vector components $k_x$ and $k_y$ are measured with respect to $\pm K$ for the valley indices $\tau = \pm 1$. 
In addition, the lower (upper) graphene layer is placed in the potential $V$ ($-V$). The parameter $\Delta$ describes the asymmetry in the energy shift of the bonding and antibonding states, which arises due to the interlayer coupling $\gamma_1$.
The Pauli spin matrices are $s_i$, with $i = \{ 0,x,y,z \}$, and $s_{\pm}^{\tau} = \frac{1}{2}(s_x\pm \textrm{i}\tau s_y)$.

The parameters $\lambda_{\textrm{I}}$ describe the intrinsic SOC of the corresponding layer and sublattice atom 
($\textrm{C}_{\textrm{A1}}, \textrm{C}_{\textrm{B1}}, \textrm{C}_{\textrm{A2}}, \textrm{C}_{\textrm{B2}}$), which are modified by the hBN layer(s).
The intrinsic SOC parameters are also present in pristine BLG and on the order of 12~$\mu$eV \cite{Konschuh2012:PRB}. 
The parameters $\lambda_{\textrm{R1}}$ and $\lambda_{\textrm{R2}}$ represent the Rashba SOC of the individual graphene layers, which can be opposite in sign \cite{Zollner2021:PRB}. 
From the symmetry point of view, also other SOC parameters are allowed, but we restrict ourselves to add only $\lambda_4$, being the interlayer SOC term connecting same sublattices, and which is the most relevant one in pristine BLG \cite{Konschuh2012:PRB,Guinea2010:NJP}.
Since we are dealing with hBN/BLG structures, the sublattice atoms are affected differently and we introduce $\lambda_4^\textrm{A}$ and $\lambda_4^\textrm{B}$.

To capture doping effects from the calculations, we introduce another parameter $E_D$, which leads to an energy shift on the model band structure and we call it the Dirac point energy. However, since the Fermi level is within the band gap for all our considered heterostructures, no charge transfer occurs and we set $E_D = 0$.
To extract the fit parameters form the DFT, we employ a least-squares routine, taking into account band energies, splittings, and spin expectation values. First, we extract the orbital parameters, $\gamma_j$, $\Delta$, and $V$. Once they are fixed, we extract the spin-orbit parameters.

\section{Band Structures, Fit Results, and Gate Tunability}
\label{sec:results}

\subsection{Pristine BLG with electric field}

\begin{figure*}[!htb]
 \includegraphics[width=.93\textwidth]{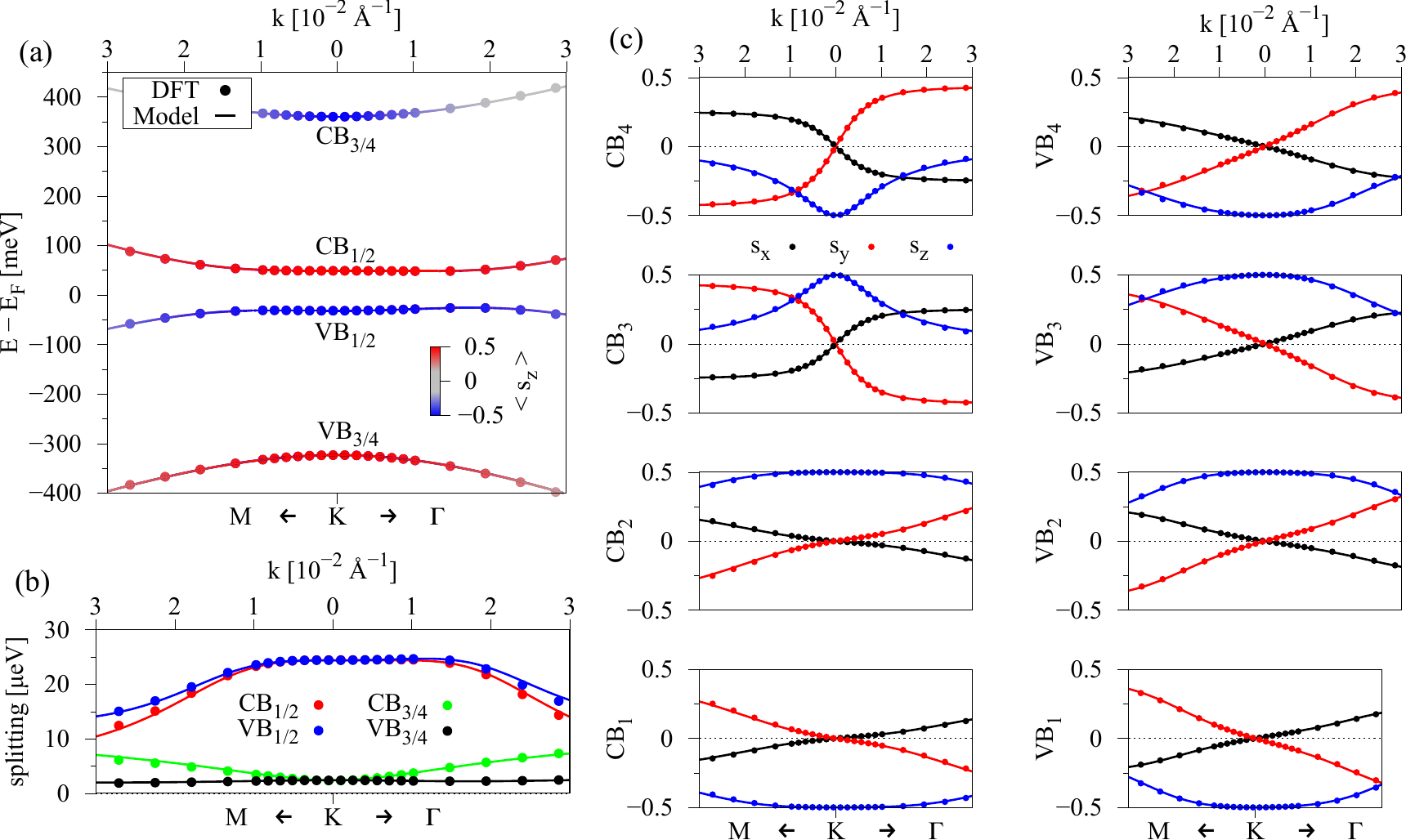}
 \caption{Calculated band properties of pristine BLG with a transverse electric field of 1~V/nm.
 (a) First-principles band structure (symbols) with a fit to the model Hamiltonian (solid lines). The symbols and lines are color coded by their $s_z$ spin expectation value. We identify the 8 relevant BLG bands by VB$_{1-4}$ (valence bands) and CB$_{1-4}$ (conduction bands). 
 (b) The splitting of the energy bands  CB$_{1/2}$ (red), VB$_{1/2}$ (blue), CB$_{3/4}$ (green), VB$_{3/4}$ (black) close to the K point and calculated model results. 
 (c) The spin expectation values of the 8 BLG bands with a comparison to the model results. 
 The fit parameters are given in Tab.~\ref{Tab:Fit_Results_pristine_BLG}. 
 }\label{Fig:pristine_BLG_efield}
\end{figure*}

\begin{table*}[htb]
\caption{\label{Tab:Fit_Results_pristine_BLG} The fit parameters of the model Hamiltonian
$\mathcal{H}$ for pristine BLG with external electric field. Additionally, we list the DFT-calculated band gap.}
\begin{ruledtabular}
\begin{tabular}{l c c c c c c c c c c }
Field [V/nm] & 0 & 0.1 & 0.25 & 0.5 & 0.75 & 1.0 & 1.25 & 1.5 & 1.75 & 2.0 \\
\hline 
$\gamma_0$ [eV] & 2.567 & 2.567 & 2.565 & 2.563 & 2.565 & 2.565 & 2.566 & 2.567 & 2.578 & 2.569 \\
$\gamma_1$ [eV] & 0.339 & 0.339 & 0.339 & 0.339 & 0.339 & 0.339 & 0.339 & 0.339 & 0.340 & 0.340   \\
$\gamma_3$ [eV] & 0.274 & 0.277 & 0.277 & 0.277 & 0.276 & 0.276 & 0.276 & 0.278 & 0.280 &  0.283  \\
$\gamma_4$ [eV] & -0.153 & -0.153 & -0.154 & -0.154 & -0.154 & -0.154 & -0.154 & -0.153 & -0.153 &  -0.153  \\
gap [meV] & 0 & 1.290 & 5.146 & 25.613 & 50.482 & 73.813 & 95.481 & 115.323 & 133.274 & 149.361 \\
$V$ [meV] & 0 & 0.959 & 2.953 & 13.640 & 27.009 & 40.334 & 53.595 & 66.768 & 79.830 & 92.768   \\
$\Delta$ [meV] & 9.783 & 9.783 & 9.783 & 9.783 & 9.782 & 9.780 & 9.776 & 9.772 & 9.765 &  9.760  \\
$\lambda_{\textrm{R1}}$ [$\mu$eV] & -6.7 & -6.9 & -7.4 & -8.4 & -9.2 & -10.6 & -11.7 & -12.9 & -13.8 & -14.8   \\
$\lambda_{\textrm{R2}}$ [$\mu$eV] & 6.7 & 6.4 & 6.1 & 5.1 & 3.7 & 2.9 & 1.8 & 0.9 & -0.3 &  -1.5 \\
$\lambda_{\textrm{I}}^\textrm{A1}$ [$\mu$eV] & 10.2 & 10.2 & 10.2 & 10.2 & 10.2 & 10.2 & 10.2  & 10.2 &  10.2 & 10.2  \\
$\lambda_{\textrm{I}}^\textrm{B1}$ [$\mu$eV] & 12.2 & 12.2 & 12.2 & 12.2 & 12.2 & 12.2 & 12.2  & 12.2 &  12.2 & 12.2 \\
$\lambda_{\textrm{I}}^\textrm{A2}$ [$\mu$eV] & 12.2 & 12.2 & 12.2 & 12.2 &  12.2 & 12.2 & 12.2  & 12.2 &  12.2 & 12.2  \\
$\lambda_{\textrm{I}}^\textrm{B2}$ [$\mu$eV] & 10.2 & 10.2 & 10.2 & 10.2 & 10.2 & 10.2 & 10.2  & 10.2 &  10.2 & 10.2  \\
$\lambda_4^\textrm{A}$ [$\mu$eV] & 12.4 & 12.4 & 13.1 & 13.0 & 13.5 & 13.7 & 14.0  & 14.1 & 14.5 & 15.0  \\
$\lambda_4^\textrm{B}$ [$\mu$eV] & 12.4 & 12.4 & 12.9 & 12.4 & 12.1 & 11.9 & 11.6  & 11.5 & 11.2 &  10.9 \\

\end{tabular}
\end{ruledtabular}
\end{table*}

The dispersion of pristine BLG features 8 parabolic bands in the vicinity of the $K$ point \cite{Konschuh2012:PRB}. In addition, the BLG bands are formed by different layers and sublattices. 
Due to inversion symmetry of the AB-stacked BLG, bands are spin degenerate without external electric field.
When a transverse electric field is applied, the two graphene layers are placed in different potentials, lifting the inversion symmetry along with spin degeneracies, and a band gap opens. Recent transport measurements have shown ultraclean band gaps in BLG heterostructures when hBN is employed as encapsulation material~\cite{Icking2022:AEM}. They also demonstrated that the gap can be tuned linearly with the applied electric field, in agreement with theory predictions~\cite{Konschuh2012:PRB}. Before we discuss the BLG/hBN heterostructures, we briefly reconsider pristine BLG in a transverse electric field to get a qualitative feeling for orbital and spin-orbit parameters, as well as the tunability of the band gap, closely following Ref.~\cite{Konschuh2012:PRB}. 

In Fig.~\ref{Fig:pristine_BLG_efield}, we show the calculated band properties of pristine BLG in the vicinity of the $K$ point and with an external transverse electric field of 1~V/nm. The first-principles data are perfectly reproduced by the parameters in Table~\ref{Tab:Fit_Results_pristine_BLG}, demonstrating the accuracy of the employed model.
In contrast to Ref.~\cite{Konschuh2012:PRB} we have even neglected some SOC parameters in the model, that are apparently irrelevant for reproducing the DFT results.
Note that the fitting procedure is not a straightforward task, since two different energy scales are present. We have hundreds of meV from the orbital parameters and few to tens of $\mu$eV from the SOC ones \cite{Konschuh2012:PRB}.
Focusing on band energies only, the orbital parameters were determined first employing the least-squares fitting routine. Then, we fixed the intrinsic SOC parameters for $\beta_1$ and $\alpha_2$ ($\alpha_1$ and $\beta_2$) to about $12~\mu$eV ($10~\mu$eV), according to Ref.~\cite{Konschuh2012:PRB}.
In other words, the dimer atoms have an intrinsic SOC value that is reduced by $2~\mu$eV, compared to the non-dimer atom. Finally, we fitted for the remaining SOC parameters.

Coming back to the band structure, see Fig.~\ref{Fig:pristine_BLG_efield}(a), we find a band gap of about 80~meV, with corresponds to about $2V$.
For the applied positive electric field, the lower graphene layer will be in a higher potential than the upper layer, fixing the sign of parameter $V$. More specifically, the low-energy bands are formed by the non-dimer atoms in the vicinity of the $K$ point, and for the positive elecric field the bands CB$_{1/2}$ (VB$_{1/2}$) are formed by atom $\beta_1$ ($\alpha_2$)~\cite{Konschuh2012:PRB,McCann2013:RPP}. The high-energy bands are formed by a combination of dimer atoms, which are split-off to about $\pm 350$~meV away from the Fermi level by the direct interlayer coupling $\gamma_1$. At the $K$ point, the low-energy bands show a spin-splitting of about $24~\mu$eV corresponding to pristine graphene~\cite{Konschuh2010:PRB,Gmitra2009:PRB,Konschuh2012:PRB}, while the high-energy band splittings are much smaller, see Fig.~\ref{Fig:pristine_BLG_efield}(b). 

At zero external field, the electrons in one layer feel an effective field due to the presence of the other layer, leading to Rashba SOCs of opposite signs in the two layers. Therefore, without external field, we cannot determine the value of the intrinsic Rashba parameters, since their effect on the bands cancel out.
However, since an external electric field is applied the Rashba effect is activated, creating vortex-like spin textures near the $K$ point.
Once an external electric field is applied, it adds to the opposing internal ones, the Rashba couplings are modified, and we can fix their values by fitting to the DFT calculation. 
For the electric field of 1~V/nm, we find the Rashba SOC parameters to be $-10.6$ and $2.9~\mu$eV.
The Rashba SOC parameters that we can extrapolate to zero electric field, by averaging the fitted parameters at 1~V/nm, are $\mp 6.7~\mu$eV for the lower/upper graphene layer. 
This value is reasonable following the argumentation in Ref.~\cite{Konschuh2012:PRB}, but does not fully agree with the value found there. 
However, it is of the same order of magnitude as in graphene/hBN heterostructures~\cite{Zollner2019:PRB}, giving us confidence in the extracted value.

\begin{figure}[!htb]
 \includegraphics[width=.99\columnwidth]{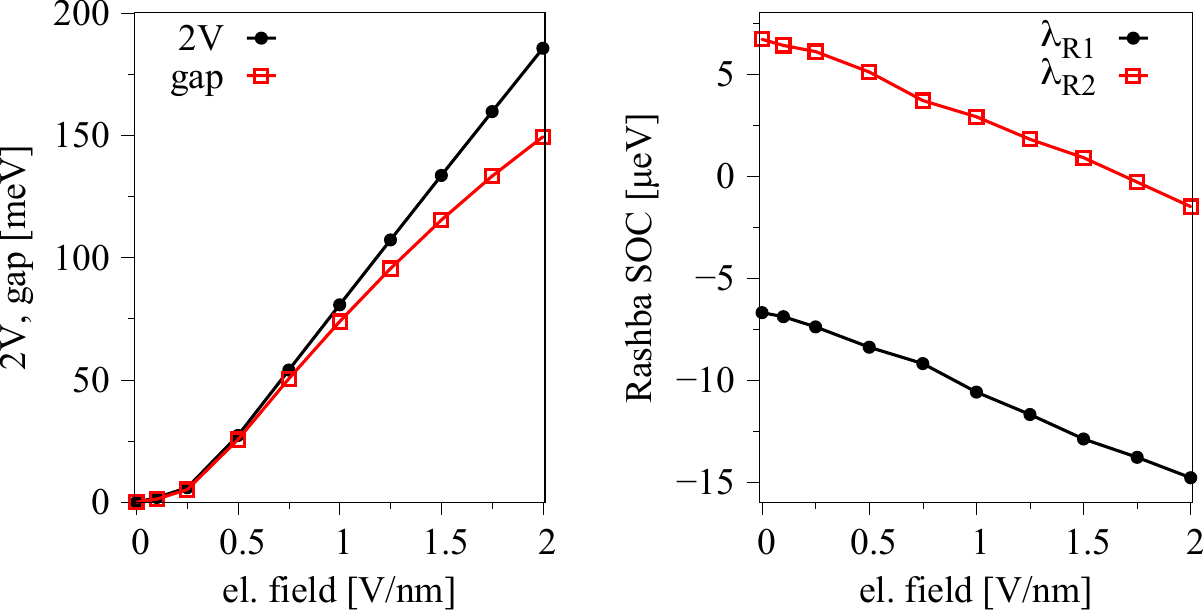}
 \caption{Evolution of the potential $V$, the band gap, and the two Rashba parameters $\lambda_{\textrm{R1}}$ and $\lambda_{\textrm{R2}}$ as function of the applied electric field for pristine BLG. The data are listed in Tab.~\ref{Tab:Fit_Results_pristine_BLG}.
 }\label{Fig:el_field_pristine_BLG}
\end{figure}

For completeness, we also summarize the fit parameters of pristine BLG without external electric field in Table~\ref{Tab:Fit_Results_pristine_BLG}. As we can see, the external field mainly influences the potential parameter $V$ and the Rashba couplings. 
In Fig.~\ref{Fig:el_field_pristine_BLG}, we show the evolution of the potential $V$, the band gap,  and the two Rashba parameters $\lambda_{\textrm{R1}}$ and $\lambda_{\textrm{R2}}$ as function of the applied electric field. We limit ourselves to show only these parameters since the others are barely affected by the field.
Both Rashba couplings follow a linear trend as expected \cite{Konschuh2012:PRB,Gmitra2009:PRB}, with a tunability of about $4~\mu$eV per V/nm. 
The potential parameter $V$ shows a nonlinear onset for small electric fields and a linear tunability for higher electric fields. For small electric fields the band gap also closely follows $2V$ and starts to drastically deviate for fields above 1~V/nm.

The above investigation will be helpful for analyzing hBN/BLG heterostructures, since the hBN layer(s) modify the overall electrostatics and the spin-orbit parameters via proximity effect~\cite{Zollner2019:PRB}.

\subsection{BLG on hBN}

\begin{figure*}[!htb]
 \includegraphics[width=.93\textwidth]{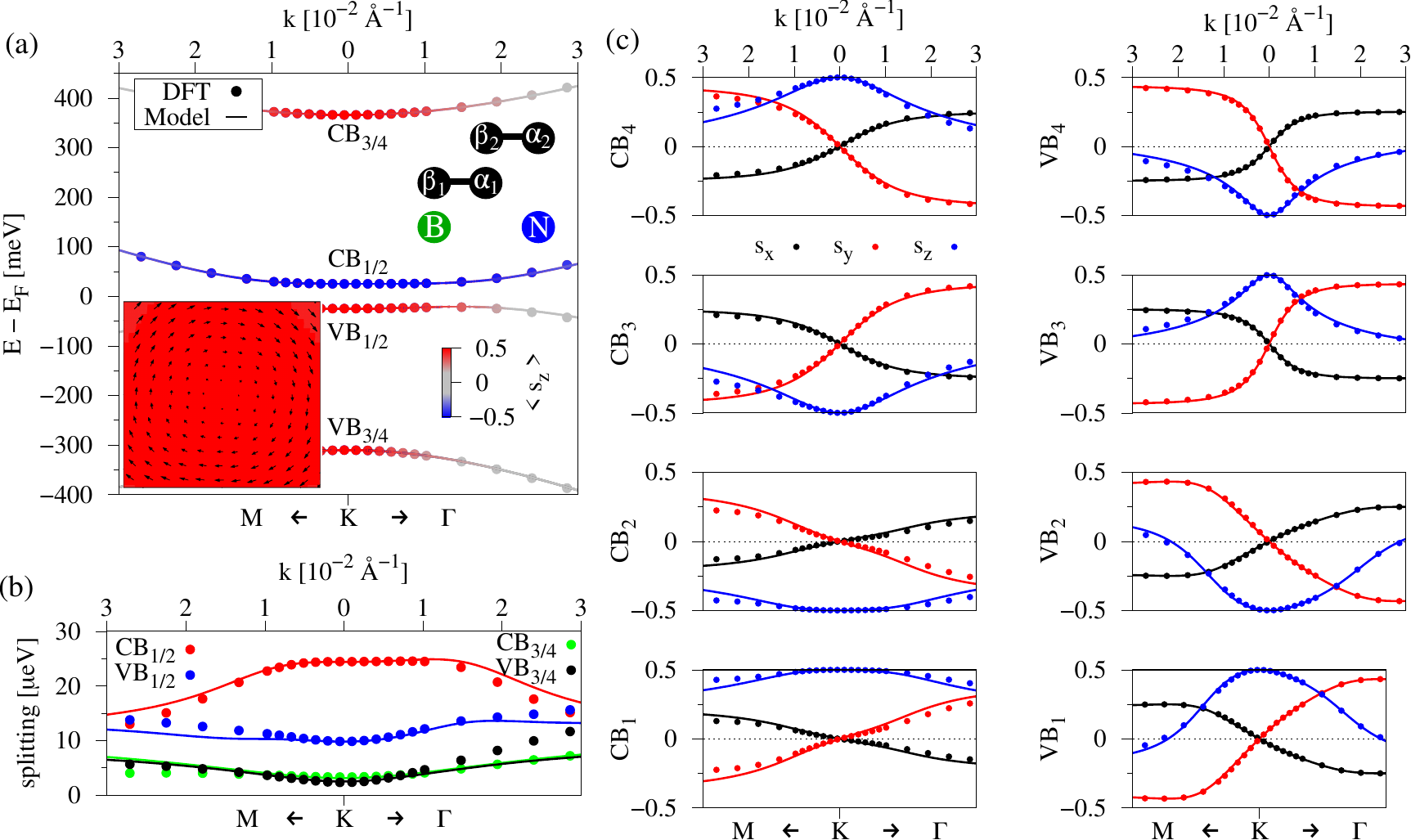}
 \caption{Calculated band properties of the BLG/hBN heterostructure in the (B$\beta_1$h$_2$,H$\alpha_1\beta_2$) stacking without electric field. Subfigures have the same meaning as in Fig.~\ref{Fig:pristine_BLG_efield}.
 The insets in (a) show the stacking sequence, as well as an exemplary spin-orbit field of VB$_1$ in the vicinity of the $K$ point.
 }\label{Fig:Fitted_DFT_BCC}
\end{figure*}

\begin{figure*}[!htb]
\includegraphics[width=.93\textwidth]{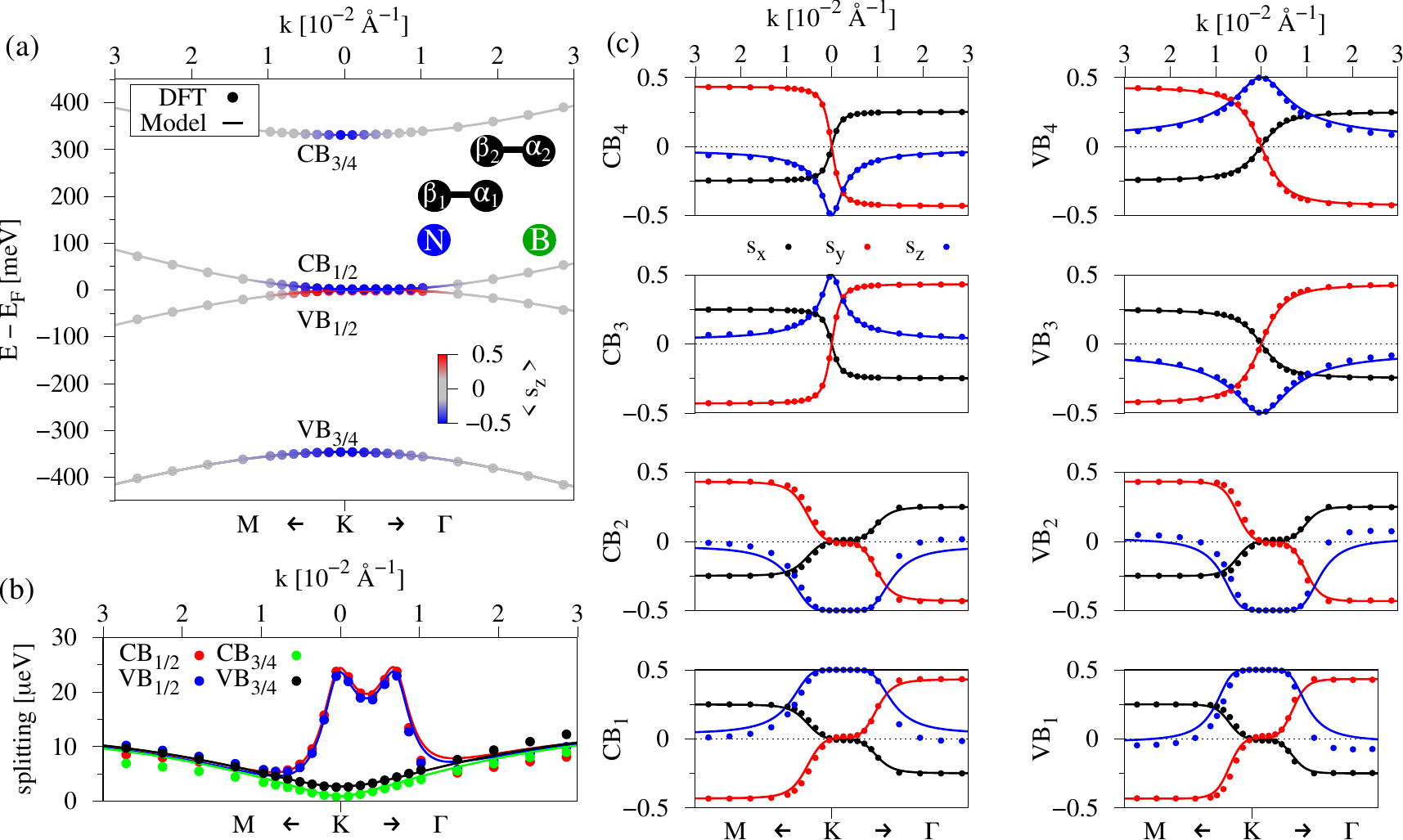}
 \caption{Calculated band properties of the BLG/hBN heterostructure in the (N$\beta_1$h$_2$,H$\alpha_1\beta_2$) stacking without electric field. Subfigures have the same meaning as in Fig.~\ref{Fig:pristine_BLG_efield}.  The inset in (a) shows the stacking sequence.
 }\label{Fig:Fitted_DFT_NCC}
\end{figure*}

\begin{figure*}[!htb]
 \includegraphics[width=.93\textwidth]{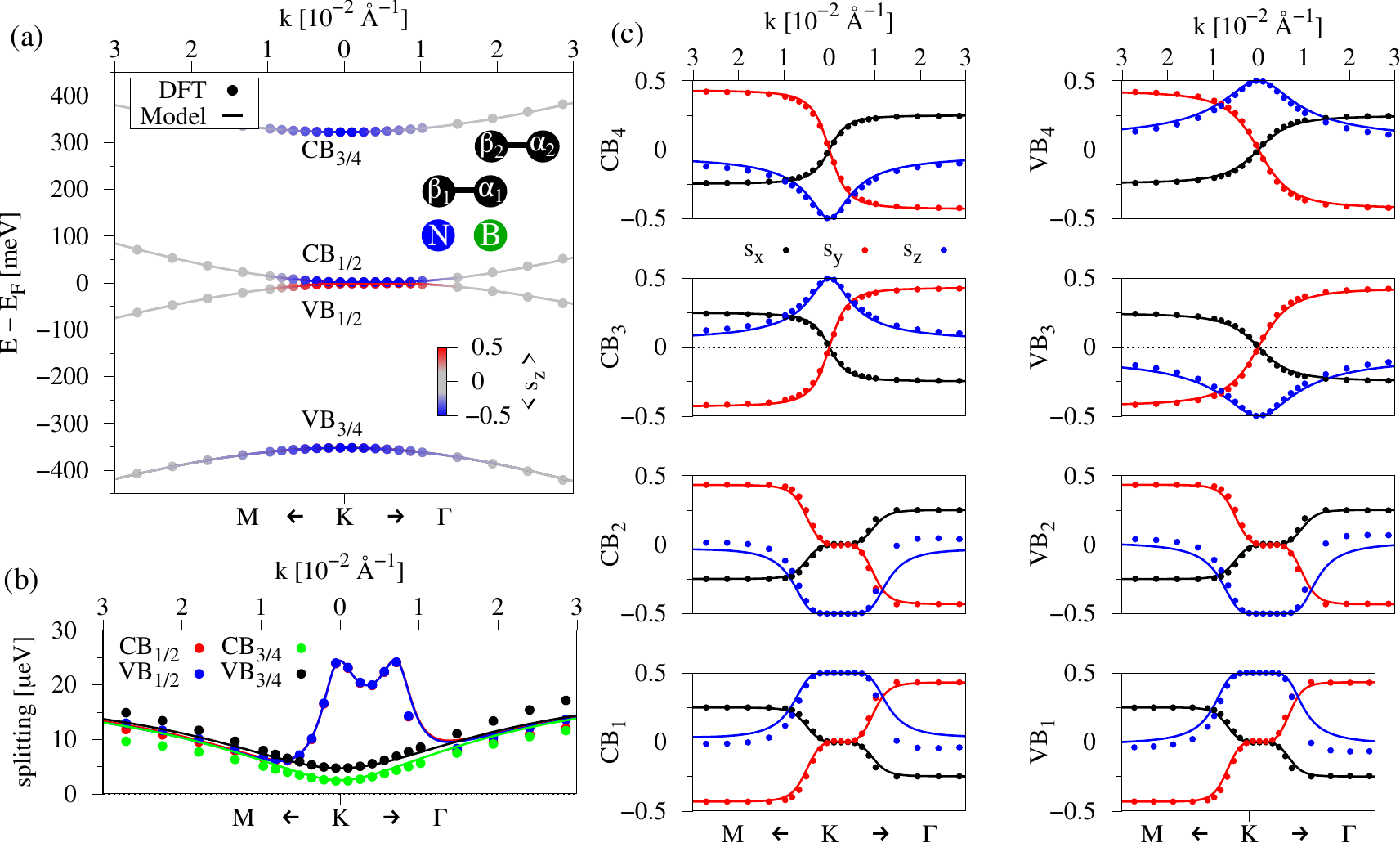}
 \caption{Calculated band properties of the BLG/hBN heterostructure in the (N$\beta_1$h$_2$,B$\alpha_1\beta_2$) stacking without electric field. Subfigures have the same meaning as in Fig.~2. The inset in (a) shows the stacking sequence.
 }\label{Fig:Fitted_DFT_BNCC}
\end{figure*}

\begin{table*}[htb]
\caption{\label{Tab:Fit_Results} The fit parameters of the model Hamiltonian
$\mathcal{H}$ for the BLG/hBN heterostructures in different stackings with and without electric field.}
\begin{ruledtabular}
\begin{tabular}{l c c c c c c }
system  & \multicolumn{2}{c}{(B$\beta_1$h$_2$,H$\alpha_1\beta_2$)} & \multicolumn{2}{c}{(N$\beta_1$h$_2$, H$\alpha_1\beta_2$)} & \multicolumn{2}{c}{(N$\beta_1$h$_2$,B$\alpha_1\beta_2$)} \\
el. field [V/nm] &  0 & 1.0 & 0 & 1.0 & 0 & 1.0 \\
\hline 
$\gamma_0$ [eV] &  2.556 & 2.544 & 2.548 & 2.525 & 2.520 & 2.543 \\
$\gamma_1$ [eV] &  0.337 & 0.339 & 0.339 & 0.337 & 0.337 & 0.335 \\
$\gamma_3$ [eV] &  0.271 & 0.272 & 0.271 & 0.276 & 0.277 & 0.279 \\
$\gamma_4$ [eV] &  -0.153 & -0.155 & -0.150 & -0.157 & -0.149 & -0.158 \\
$V$ [meV] &  -25.087 & 5.906 & -1.291 & 37.361 & -1.440 & 36.543 \\
$\Delta$ [meV] &  27.584 & 26.698  & -7.895 & -9.461 & -15.319 & -16.201 \\
$\lambda_{\textrm{R1}}$ [$\mu$eV] &  4.1 & 0.1 & 6.0 & 2.0 & 9.2 & 5.2 \\
$\lambda_{\textrm{R2}}$ [$\mu$eV] &  6.8 & 2.9 &  10.1 & 6.1 & 10.9 & 7.0 \\
$\lambda_{\textrm{I}}^\textrm{A1}$ [$\mu$eV] & 7.4 & 7.4 & 12.0 & 12.0 & 13.8 & 13.8 \\
$\lambda_{\textrm{I}}^\textrm{B1}$ [$\mu$eV] &  4.9 & 4.9 & 11.8 & 11.8 & 12.2 & 12.2 \\
$\lambda_{\textrm{I}}^\textrm{A2}$ [$\mu$eV] &  12.2 & 12.2 & 12.2 & 12.2 & 12.2 & 12.2 \\
$\lambda_{\textrm{I}}^\textrm{B2}$ [$\mu$eV] &  10.2 & 10.2 & 10.2 & 10.2 & 10.2 & 10.2 \\
$\lambda_4^\textrm{A}$ [$\mu$eV] &  24.4 & 21.3 & 17.3 & 14.8 & 22.2 & 13.3 \\
$\lambda_4^\textrm{B}$ [$\mu$eV] &  22.2 & 18.7 & 19.9 & 11.0 & 18.6 & 11.3 \\
\end{tabular}
\end{ruledtabular}
\end{table*}

In the case of hBN/BLG structures, we follow a similar fitting strategy as before. Focusing on band energies only, the orbital parameters are determined first. Then, we fix the intrinsic SOC parameters, combining the knowledge about pristine graphene~\cite{Gmitra2009:PRB}, pristine BLG \cite{Konschuh2012:PRB}, and graphene/hBN heterostructures \cite{Zollner2019:PRB}. 
For example, in the (B$\beta_1$h$_2$,H$\alpha_1\beta_2$) stacking, and assuming that the hBN layer has no impact on the top graphene layer, we fix the intrinsic SOC parameter for $\alpha_2$ ($\beta_2$) to about $12~\mu$eV ($10~\mu$eV), according to Ref.~\cite{Konschuh2012:PRB}, and with the knowledge about the $2~\mu$eV difference between dimer/non-dimer atoms.
The intrinsic SOC parameters of the lower graphene layer are proximity modified by the hBN \cite{Zollner2019:PRB}. For this particular graphene/hBN stacking, the atom $\beta_1$ ($\alpha_1$) would have an intrinsic SOC of about $5.0~\mu$eV ($9.4~\mu$eV). Since $\alpha_1$ is still a dimer atom within BLG, we have to reduce the value additionally by $2~\mu$eV, while $\beta_1$ is fixed by the previous consideration. 
Finally, we fit for the remaining SOC parameters that best reproduce the band splittings and spin expectation values, taking also into account the previously determined intrinsic Rashba SOC of pristine BLG and the Rashba SOC in graphene/hBN heterostructures \cite{Zollner2019:PRB}. 
For example, the lower graphene layer in the (B$\beta_1$h$_2$,H$\alpha_1\beta_2$) stacking, acquires a Rashba SOC of $-6.7~\mu$eV ($10.7~\mu$eV) from the upper graphene (lower hBN) layer, resulting in $\lambda_{\textrm{R1}} = 4.1~\mu$eV.
Following this procedure, we can quite accurately reproduce the dispersion, spin splittings, and spin expectation values of the three hBN/BLG heterostructure stackings we consider, see Fig.~\ref{Fig:Fitted_DFT_BCC}, Fig.~\ref{Fig:Fitted_DFT_NCC}, and Fig.~\ref{Fig:Fitted_DFT_BNCC}. 
The model parameters are summarized in Table~\ref{Tab:Fit_Results}.  
The remaining discrepancies between the model and the DFT results arise due to the influence of hBN $p$ orbitals, disturbing the effective $p_z$-orbital-based model description of BLG, similar as in graphene/hBN heterostructures~\cite{Zollner2019:PRB}.

Focusing on the energetically most favorable (B$\beta_1$h$_2$,H$\alpha_1\beta_2$) stacking, see Fig.~\ref{Fig:Fitted_DFT_BCC}, we find a sizable orbital gap ($\sim 50$~meV) at the Fermi level without external electric field. Remember that the layer and sublattice character plays an important role in the BLG dispersion \cite{Gmitra2017:PRL,Zollner2020:PRL,Zollner2021:PRB}. The hBN layer introduces a significant potential difference between the graphene layers, opening the band gap. 
For this specific stacking, the low-energy valence bands, VB$_{1/2}$, are formed by the bottom graphene layer, being in direct contact to hBN. More specifically, it is the non-dimer atom $\beta_1$ forming these bands. In contrast, the non-dimer atom $\alpha_2$ forms the low-energy conduction bands, CB$_{1/2}$. The low-energy valence (conduction) band is spin split by about $10~\mu$eV ($24~\mu$eV).
The reason for this difference in the splitting is due to short-range proximity effect. The intrinsic SOC parameter of $\beta_1$ ($\approx 5~\mu$eV) is proximity-modified due to the subjacent hBN layer, while the intrinsic SOC parameter of $\alpha_2$ ($\approx 12~\mu$eV) stays intact.
The vortex-like spin-texture of the low-energy bands, which is due to the Rashba SOC, is visualized in the inset of Fig.~\ref{Fig:Fitted_DFT_BCC}(a) as an exemplary case.

The high-energy bands, being far away from the Fermi level, are formed by dimer atoms $\alpha_1$ and $\beta_2$. Moreover, the asymmetry $\Delta$ is strongly affected by the hBN layer, see Table~\ref{Tab:Fit_Results}. 

The low-energy dispersions and spin splittings of the other stackings, see Fig.~\ref{Fig:Fitted_DFT_NCC} for (N$\beta_1$h$_2$,H$\alpha_1\beta_2$) and Fig.~\ref{Fig:Fitted_DFT_BNCC} for (N$\beta_1$h$_2$,B$\alpha_1\beta_2$), are markedly different to the (B$\beta_1$h$_2$,H$\alpha_1\beta_2$) stacking. First, we find that the orbital gap is nearly closed. This cannot be seen on the chosen energy scale, but the gap is about $2V$, as we have found above. In addition, the low-energy band splittings are nearly equal. The reason is that the hBN layer, for these two stackings, does not alter the intrinsic SOC parameter of $\beta_1$ as much as before. 
In conclusion, the low-energy band properties of BLG are mainly influenced by hBN in terms of the orbital band gap, while SOC parameters stay on the order of few to tens of $\mu$eV. The atomic stacking configuration strongly influences the size of the band gap, which can range from few to tens of meV.

Applying an external electric field essentially adds to the internal field, that arises due the asymmetry introduced by the hBN layer. For example, in the (B$\beta_1$h$_2$,H$\alpha_1\beta_2$) stacking, an external electric field of 1~V/nm tunes the parameter $V$ from about $-26$ to $6$~meV. In other words, the band character of the low-energy bands has switched, and now the valence (conduction) bands are formed by $\alpha_2$ ($\beta_1$). In addition, since the intrinsic SOC parameters of these atoms do not change with applied field, the band splittings are also switched accordingly. This electric field induced switching of the low-energy spin splittings resembles the spin-orbit and exchange valve effects as in Refs.~\cite{Gmitra2017:PRL,Zollner2018:NJP,Zollner2021:PRB} for BLG heterostructures. For a certain external field value, one can even fully counter the internal field and close the gap. 
In addition, the Rashba SOC parameters are affected by the field, which are roughly tunable by $4~\mu$eV per V/nm. Otherwise, all other orbital and spin-orbit parameters are barely affected by the external electric field. 
The model parameters with electric field are also summarized in Table~\ref{Tab:Fit_Results}.

\subsection{hBN encapsulated BLG}
In a similar way as for the hBN/BLG heterostructures, we fit the band properties of the hBN encapsulated BLG structures. First, we determine the orbital parameters from the band energies only. 
The intrinsic SOC parameters are again fixed according to Ref.~\cite{Zollner2019:PRB} for the different stacking configurations of the individual top and bottom graphene/hBN bilayers within the heterostructures and from the dimer/non-dimer difference discussed above. Then, we fit for the remaining SOC parameters that best reproduce the band splittings and spin expectation values, again taking into account the argumentation for the Rashba SOCs. The fit results are summarized in Table~\ref{Tab:Fit_Results_encapsulated}. 

\begin{table*}[htb]
\caption{\label{Tab:Fit_Results_encapsulated} The fit parameters of the model Hamiltonian $\mathcal{H}$ for the hBN encapsulated BLG heterostructures in different stackings and with different numbers of hBN layers. Numbers without (with) brackets correspond to BLG encapsulated by monolayers of hBN (top monolayer hBN and bottom bilayer hBN).}
\begin{ruledtabular}
\begin{tabular}{l c c c c c c}
system & (B$\beta_1$h$_2$N,H$\alpha_1\beta_2$H) & 
(N$\beta_1$h$_2$B,H$\alpha_1\beta_2$H) & 
(B$\beta_1$h$_2$B,H$\alpha_1\beta_2$H) & 
(B$\beta_1$h$_2$H,H$\alpha_1\beta_2$B) & 
(N$\beta_1$h$_2$H,H$\alpha_1\beta_2$B) & 
(N$\beta_1$h$_2$H,B$\alpha_1\beta_2$B) \\
\hline 
$\gamma_0$ [eV] & 2.563 (2.561) & 2.527 (2.526) & 2.570 (2.565) & 2.567 (2.562) & 2.517 (2.511) & 2.506 (2.505)  \\
$\gamma_1$ [eV] & 0.337 (0.337) & 0.337 (0.337) & 0.337 (0.336) & 0.335 (0.335) & 0.335 (0.335) & 0.334 (0.333)  \\
$\gamma_3$ [eV] & 0.266 (0.266) & 0.271 (0.273) & 0.269 (0.272) & 0.268 (0.273) & 0.273 (0.276) & 0.273 (0.274)  \\
$\gamma_4$ [eV] & -0.151 (-0.152) & -0.150 (-0.150) & -0.149 (-0.150) & -0.148 (-0.150) & -0.149 (-0.149) & -0.149 (-0.150)    \\
$V$ [meV] &  0 (-1.185) & 0 (-1.350) & -17.980 (-24.685) & -17.215 (-23.853)  & 0.146 (-1.178) & 0 (-1.319)  \\
$\Delta$ [meV] & 45.137 (45.020) & -25.810 (-24.997) & 9.649 (9.517)  & 2.285 (2.186) & -33.176 (-32.334) & -40.526 (-39.781)  \\
$\lambda_{\textrm{R1}}$ [$\mu$eV] & 4.0 (4.4) & 6.0 (6.6) & 2.5 (2.9) & 1.5 (2.0) & 2.5 (2.9) &  10.2 (10.2) \\
$\lambda_{\textrm{R2}}$ [$\mu$eV] & -4.0 (-3.6) & -6.0 (-6.1) & -6.6 (-5.6) & -9.3 (-8.4) & -7.6 (-7.2) & -10.2 (-10.3)\\
$\lambda_{\textrm{I}}^\textrm{A1}$ [$\mu$eV] & 7.4 (7.4) & 12.0 (12.0) & 7.4 (7.4) & 7.4 (7.4) & 12.0 (12.0) & 13.8 (13.8)\\
$\lambda_{\textrm{I}}^\textrm{B1}$ [$\mu$eV] & 4.9 (4.9) & 11.8 (11.8) & 4.9 (4.9) & 4.9 (4.9) & 11.8 (11.8) & 12.2 (12.2)\\
$\lambda_{\textrm{I}}^\textrm{A2}$ [$\mu$eV] & 4.9 (4.9) & 11.8 (11.8) & 11.8 (11.8) & 12.2 (12.2) & 12.2 (12.2) & 12.2 (12.2)\\
$\lambda_{\textrm{I}}^\textrm{B2}$ [$\mu$eV] & 7.4 (7.4) & 12.0 (12.0) & 12.0 (12.0) & 13.8 (13.8) & 13.8 (13.8) & 13.8 (13.8) \\
$\lambda_4^\textrm{A}$ [$\mu$eV] & 22.0 (21.9) & 11.8 (14.3) & 12.0 (11.8) & 11.0 (11.6) & 24.2 (20.7) & 15.0 (13.8) \\
$\lambda_4^\textrm{B}$ [$\mu$eV] & 22.0 (22.5) & 11.8 (15.5) & 17.4 (17.6) & 17.2 (17.4) & 27.2 (23.7) & 15.0 (14.0) \\
\end{tabular}
\end{ruledtabular}
\end{table*}

\begin{figure*}[!htb]
 \includegraphics[width=.98\textwidth]{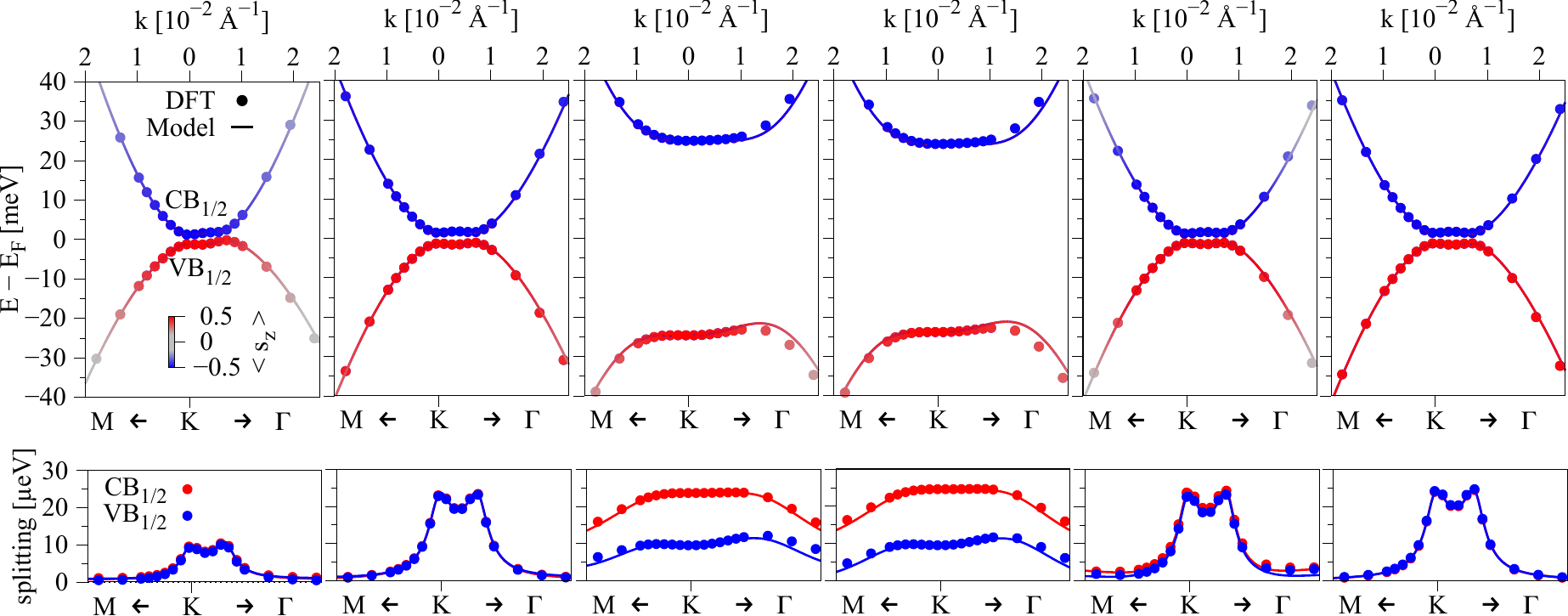}
 \caption{Calculated low-energy bands (top) and corresponding spin splittings (bottom) of BLG encapsulated by bottom bilayer hBN and top monolayer hBN. From left to right, the subfigures correspond to the stackings as in Table~\ref{Tab:Fit_Results_encapsulated}.
 }\label{Fig:LE_hBN_hBN_BLG_hBN}
\end{figure*}

For the hBN encapsulated BLG heterostructures, inversion symmetry can be restored for special stacking configurations. For example, in the (B$\beta_1$h$_2$N,H$\alpha_1\beta_2$H) stacking and considering surrounding hBN monolayers, we find that the potential difference $V=0$, just as for pristine BLG without external electric field. In contrast to pristine BLG, the asymmetry $\Delta$ is strongly modified due to the hBN layers. Moreover, also the SOC parameters are heavily influenced. In particular, the Rashba SOCs have the opposite sign and nearly the same magnitude compared to pristine BLG. The intrinsic SOC parameters are again symmetric ($\lambda_{\textrm{I}}^\textrm{A1} = \lambda_{\textrm{I}}^\textrm{B2}$, $\lambda_{\textrm{I}}^\textrm{B1} = \lambda_{\textrm{I}}^\textrm{A2}$), but reduced in value due to the hBN layers. Remarkably, the parameters $\lambda_4$ are significantly enhanced compared to pristine BLG. Similarly, the (N$\beta_1$h$_2$B,H$\alpha_1\beta_2$H) and (N$\beta_1$h$_2$H,B$\alpha_1\beta_2$B) stacking configurations also have inversion symmetry and the band gap closes. 

Once we employ top monolayer hBN and bottom bilayer hBN for the encapsulation, inversion symmetry is again broken for all stackings, and a finite $V$ arises. However, the effect of adding a second hBN layer is not dramatic anymore. For example, the (B$\beta_1$h$_2$N,H$\alpha_1\beta_2$H) stacking shows a potential difference of $V \approx -1.2$~meV due to the additional hBN layer, while SOC parameters are barely affected anymore. From that, we conclude that only the neighboring hBN layers have a strong impact on the spin physics in BLG. Nevertheless, the exact encapsulation can drastically influence the low-energy bands. 

In Fig.~\ref{Fig:LE_hBN_hBN_BLG_hBN}, we show the low-energy bands and corresponding spin splittings for the different stacking configurations with top monolayer and bottom bilayer hBN. Again, we see very good agreement of DFT and model results, giving us further confidence in our model Hamiltonian and the fitted parameters.
We find band gaps in the range of 0 to 50~meV for the different structures. Experimentally, when a moir\'{e} pattern is present, one can therefore expect local variations in the band gap, due to the different atomic stackings.

\subsection{Gate tunability}

\begin{table*}
\caption{\label{Tab:Fit_Results_encap_field} The fit parameters of the model Hamiltonian
$\mathcal{H}$ for BLG with top monolayer hBN and bottom bilayer hBN in the (B$\beta_1$h$_2$N,H$\alpha_1\beta_2$H) stacking with external electric field. Additionally, we list the DFT-calculated band gap.}
\begin{ruledtabular}
\begin{tabular}{l c c c c c c c c c c c c c}
Field [V/nm] & -2.0 & -1.5 & -1.0 & -0.75 & -0.5 & -0.25 & 0.0 & 0.25 & 0.5 & 0.75 & 1.0 & 1.5 & 2.0 \\
\hline 
$\gamma_0$ [eV] & 2.558 & 2.558 & 2.559 & 2.559 & 2.559 & 2.559 & 2.561 & 2.560 &  2.559 & 2.558 &  2.558 & 2.558 & 2.557 \\
$\gamma_1$ [eV] & 0.336 & 0.336 & 0.337 & 0.337 & 0.337 & 0.337 & 0.337 & 0.337 &  0.337 & 0.337 &  0.337 & 0.336 & 0.336 \\
$\gamma_3$ [eV] & 0.319 & 0.296 & 0.278 & 0.272 & 0.268 & 0.266 & 0.266 & 0.266 &  0.266 & 0.269 &  0.273 & 0.290 & 0.311 \\
$\gamma_4$ [eV] & -0.151 & -0.151 & -0.152 &  -0.152 &  -0.152 &  -0.152 &  -0.152 & -0.152 &  -0.152 & -0.152 &  -0.152 & -0.152 & -0.151  \\
gap [meV] & 178.752 & 144.873 & 100.845 & 74.937 &  46.745 &  17.251 & 1.362 & 2.624 &  22.831 & 52.152 &  80.038 & 128.363 & 166.330 \\
$V$ [meV] & -120.184 & -89.450 & -57.897 & -41.892 &  -25.796 &  -9.664 & -1.185 & 1.854 &  12.664 & 28.816 &  44.906 & 76.747 & 107.89 \\
$\Delta$ [meV] & 45.533 & 45.331 & 45.179 & 45.121 &  45.076 &  45.042 & 45.020 & 45.009 &  45.010 & 45.022 &  45.044 & 45.119 & 45.228 \\
$\lambda_{\textrm{R1}}$ [$\mu$eV] & 14.5 & 11.6 & 9.0  & 7.5 & 6.4 & 5.0 &  4.4 & 3.0 & 2.2 & 0.7 &  -0.8 & -4.3 & -7.7 \\
$\lambda_{\textrm{R2}}$ [$\mu$eV] & 9.8 & 6.6 & 3.1 & 1.6 &  -0.3 & -2.0 &  -3.6 & -4.2 & -5.8 & -7.3 &  -8.9 & -11.6 & -14.4 \\
$\lambda_{\textrm{I}}^\textrm{A1}$ [$\mu$eV] & 7.4 & 7.4 & 7.4 & 7.4 & 7.4 & 7.4 & 7.4 & 7.4 & 7.4 & 7.4 & 7.4 & 7.4 & 7.4 \\
$\lambda_{\textrm{I}}^\textrm{B1}$ [$\mu$eV] & 4.9 & 4.9 & 4.9 & 4.9 & 4.9 & 4.9 & 4.9 & 4.9 & 4.9 & 4.9 & 4.9 & 4.9 & 4.9 \\
$\lambda_{\textrm{I}}^\textrm{A2}$ [$\mu$eV] & 4.9 & 4.9 & 4.9 & 4.9 & 4.9 & 4.9 & 4.9 & 4.9 & 4.9 & 4.9 & 4.9 & 4.9 & 4.9 \\
$\lambda_{\textrm{I}}^\textrm{B2}$ [$\mu$eV] & 7.4 & 7.4 & 7.4 & 7.4 & 7.4 & 7.4 & 7.4 & 7.4 & 7.4 & 7.4 & 7.4 & 7.4 & 7.4 \\
$\lambda_4^\textrm{A}$ [$\mu$eV] & 25.8 & 23.9 & 22.6 & 22.4 &  22.3 & 22.4 &  21.9 & 23.2 &  23.6 & 24.3 &  25.1 & 26.4 & 28.1 \\
$\lambda_4^\textrm{B}$ [$\mu$eV] & 26.9 & 25.3 & 24.5 & 24.0 &  23.7 & 23.3 &  22.5 & 23.8 &  23.1 & 22.9 &  22.8 & 23.4 & 25.0 \\

\end{tabular}
\end{ruledtabular}
\end{table*}

In the following, we consider the energetically most favorable structure of hBN-encapsulated BLG and again study the electric field tunability. In particular, we consider the structure with top monolayer hBN and bottom bilayer hBN in the (B$\beta_1$h$_2$N,H$\alpha_1\beta_2$H) stacking configuration. The fit results as function of the external electric field are summarized in Table~\ref{Tab:Fit_Results_encap_field}.

\begin{figure}[!htb]
 \includegraphics[width=.99\columnwidth]{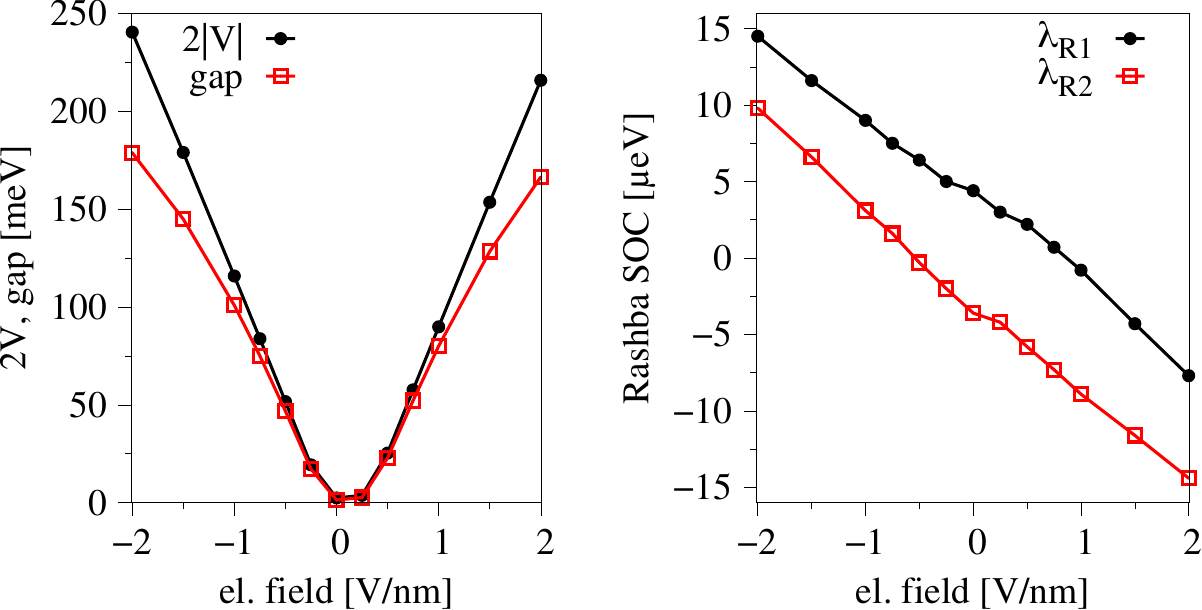}
 \caption{Evolution of the potential $V$ (absolute value), the band gap, and the two Rashba parameters $\lambda_{\textrm{R1}}$ and $\lambda_{\textrm{R2}}$ as function of the applied electric field for BLG with top monolayer hBN and bottom bilayer hBN in the (B$\beta_1$h$_2$N,H$\alpha_1\beta_2$H) stacking.
 }\label{Fig:el_field_BCCB}
\end{figure}

In Fig.~\ref{Fig:el_field_BCCB}, we show the evolution of the potential $V$, the band gap,  and the two Rashba parameters $\lambda_{\textrm{R1}}$ and $\lambda_{\textrm{R2}}$ as function of the applied electric field, similar as above for pristine BLG. 
Both Rashba couplings again follow a linear trend, with a tunability of about $5.5~\mu$eV per V/nm. This value is a bit larger compared to pristine BLG, but still of the same magnitude. 
The potential parameter $V$ shows a nonlinear onset for small electric fields, and a linear tunability for higher electric fields. For small electric fields, the band gap also closely follows $2V$ and starts to drastically deviate for fields above 1~V/nm. The asymmetry for small electric field values originates from the structural asymmetry. 

\section{Twist Angle Dependent Band Gap}
\label{sec:twist}
So far, we have considered specific stacking configurations to get a quantitative and qualitative feeling for orbital and spin-orbit properties. However, this only provides a glimpse of local regions in a generic experimental setup. 
In fact, in realistic macroscopic experimental samples, stacking faults and disorder can be present ~\cite{Warner2009:NL,Wilson2020:PRR}.
Even with the knowledge about the stacking dependent electronic spectra we have analyzed above, the interpretation of experimental data is further complicated. Here, we do not consider lateral disorder (stacking faults) within BLG, which may change electronic states.

Instead, we consider twisted Bernal-BLG/hBN heterostructures, where essentially several of the investigated stacking configurations are present simultaneously. Since these structures can become quite large and are computationally very demanding, we employ the plane wave based DFT code \texttt{Quantum Espresso}~\cite{Giannozzi2009:JPCM} and neglect the effects of SOC.
The above analysis has already shown that SOC effects are much smaller than orbital ones, making this a valid approach. 
 In any case, the plane wave and pseudopotential method, implemented in \texttt{Quantum Espresso}, cannot give correct spin-orbit splittings in graphene, since the relevant $d$-orbitals are missing \cite{Gmitra2009:PRB,Konschuh2010:PRB}. Nevertheless, on an orbital level, the dispersions are the same as calculated with \texttt{WIEN2k}, and we can safely study the electric field tunability.

The twisted heterostructures were set-up with the {\tt atomic simulation environment (ASE)} \cite{ASE} and the {\tt CellMatch} code \cite{Lazic2015:CPC}, implementing the coincidence lattice method \cite{Koda2016:JPCC,Carr2020:NRM}. We keep the lattice constant of BLG fixed at $a = 2.46$~\AA~and strain hBN less than $\pm 0.2$\%.
In particular, we consider 4 different twisted BLG/hBN structures, with twist angles between hBN and BLG of $6.42^{\circ}$ (468 atoms, $a_{\textrm{m}} = 21.865$~\AA), $10.89^{\circ}$ (166 atoms, $a_{\textrm{m}} = 13.017$~\AA), $14.56^{\circ}$ (374 atoms, $a_{\textrm{m}} = 19.526$~\AA), and $19.59^{\circ}$ (468 atoms, $a_{\textrm{m}} = 21.865$~\AA), where $a_{\textrm{m}}$ is the lattice constant of the twisted supercell. 
We use an energy cutoff for charge density of $480$~Ry and
the kinetic energy cutoff for wavefunctions is $60$~Ry for the scalar relativistic pseudopotentials with the projector augmented wave method~\cite{Kresse1999:PRB} with the Perdew-Burke-Ernzerhof exchange correlation functional~\cite{Perdew1996:PRL}. 
Self-consistent calculations are carried out with a $k$-point sampling of $12\times 12\times 1$.
Again, we employ DFT-D2 vdW corrections~\cite{Grimme2006:JCC,Grimme2010:JCP,Barone2009:JCC}.
In order to simulate quasi-2D systems, we add a vacuum of about $20$~\AA~to avoid interactions between periodic images in our slab geometry. To get proper interlayer distances and to capture possible moir\'{e} reconstructions, we allow all atoms to move freely within the heterostructure geometry during relaxation. Relaxation is performed until every component of each force is reduced below $1\times10^{-4}$~[Ry/$a_0$], where $a_0$ is the Bohr radius.

After relaxation, we calculate the mean interlayer distances, $d_{\textrm{G-G}}$ and $d_{\textrm{G-hBN}}$, and the standard deviations, $\Delta z_{\textrm{hBN}}$, from the $z$ coordinates of the C, B, and N atoms. The standard deviations represent the amount of rippling of the hBN layer. From the calculated band energies, we extract the global band gap. All these results are listed in Table~\ref{Tab:twisted_struct}.
We find relaxed interlayer distances of 3.24~\AA, between the graphene layers, and 3.26~\AA, between the graphene and hBN layers. These values are in agreement with earlier calculations~\cite{Giovannetti2007:PRB,Zollner2019:PRB}. The global band gap increases linearly with the twist angle. Fitting the twisted band gap data to a linear function, we find a $0^{\circ}$ global band gap of 23.23~meV and a slope of 0.39~meV/${^{\circ}}$. 

In general, the global band gap results from the average of the band gaps of the different local stacking configurations.
Therefore, we also consider the high-symmetry stackings from above, which correspond to $0^{\circ}$ and $60^{\circ}$ twist angles and fix the interlayer distances of 3.24~\AA, between the graphene layers, and 3.26~\AA, between the graphene and hBN layers, which are the average of the twisted structures.  
From the band gaps of the individual stackings, see Table~\ref{Tab:twisted_struct_0deg}, we can also calculate the average gap, which we determine to be 25.82~meV, in good agreement with the extrapolated one of the twisted structures.

    \begin{table}[htb]
    \caption{\label{Tab:twisted_struct} Calculated relaxed interlayer distances, hBN ripplings, and band gaps of the twisted BLG/hBN structures. }
    \begin{ruledtabular}
    \begin{tabular}{ccccc}
    $\vartheta$ [°] & $d_{\textrm{G-G}}$~[\AA] & $d_{\textrm{G-hBN}}$~[\AA] & $\Delta z_{\textrm{hBN}}$ [pm] & gap [meV] \\ \hline
6.42 & 3.2379 & 3.2637 & 3.9260 & 25.58   \\
10.89 & 3.2396 & 3.2606 & 2.3575 &  27.50\\
14.56 & 3.2385 & 3.2562 & 1.4732 &  29.16\\
19.59 & 3.2448 & 3.2608 & 0.5418 &  30.63 \\ 
    \end{tabular}
    \end{ruledtabular}
    \end{table}

    \begin{table}[htb]
    \caption{\label{Tab:twisted_struct_0deg} Calculated total energies and band gaps of the different $0^{\circ}$ stackings, with fixed interlayer distances of 3.24~\AA, between the graphene layers, and 3.26~\AA, between the graphene and hBN layers. }
    \begin{ruledtabular}
    \begin{tabular}{ccc}
     stacking  & E$_{\textrm{tot}}$-E$_0$ [meV] & gap [meV] \\ \hline
(B$\beta_1$h$_2$,H$\alpha_1\beta_2$) & 0 & 46.48\\
(N$\beta_1$h$_2$,H$\alpha_1\beta_2$) & 22.62 & 11.38 \\
(N$\beta_1$h$_2$,B$\alpha_1\beta_2$) & 30.36 & 11.77\\
(B$\beta_1$h$_2$,N$\alpha_1\beta_2$) & 29.35 & 53.43 \\
(H$\beta_1$h$_2$,N$\alpha_1\beta_2$) & 21.80 & 18.63 \\
(H$\beta_1$h$_2$,B$\alpha_1\beta_2$) & 0.15 & 13.22 \\
\hline
average & - & 25.82\\
    \end{tabular}
    \end{ruledtabular}
    \end{table}

\begin{figure*}[!htb]
 \includegraphics[width=.99\textwidth]{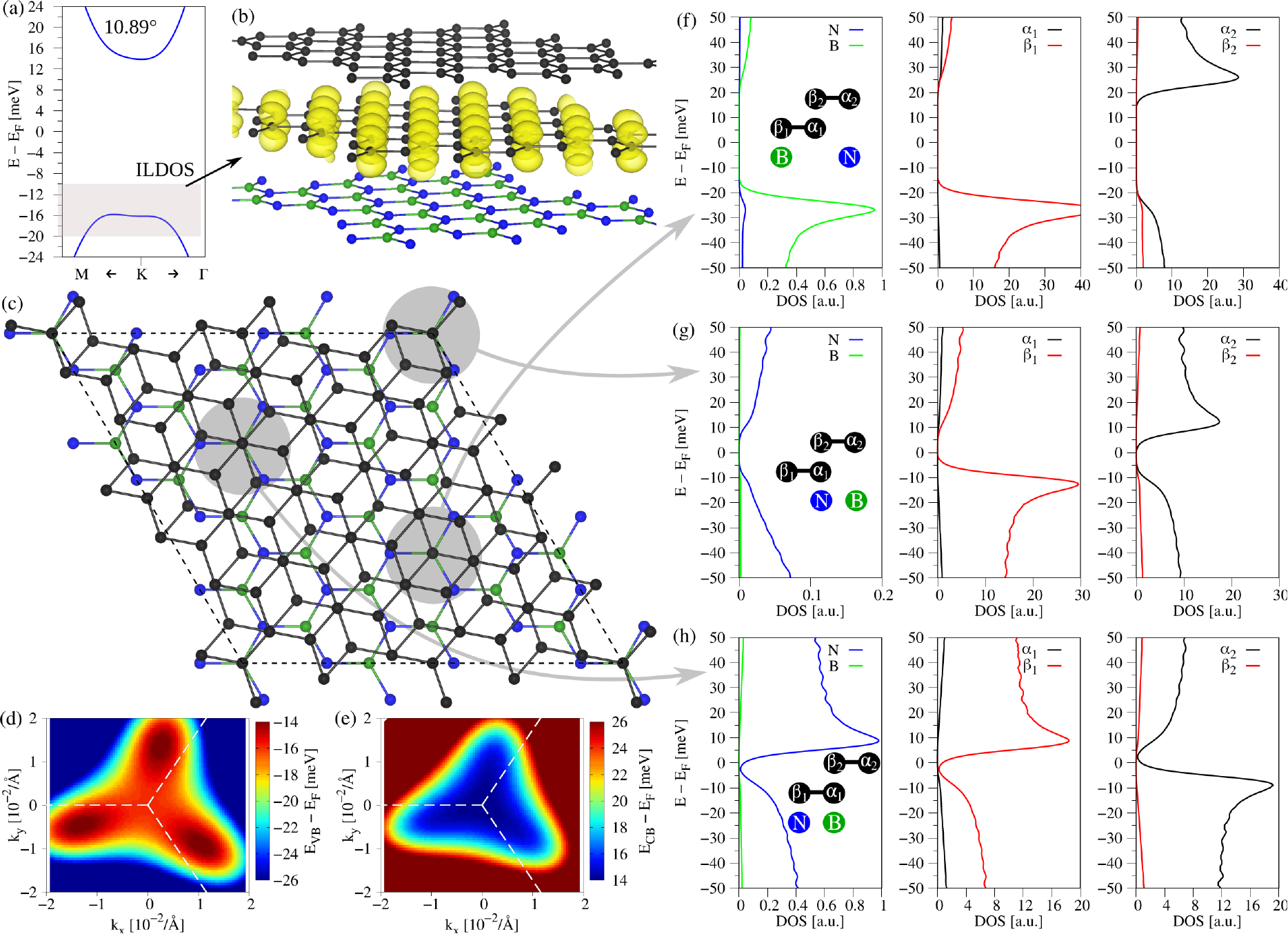}
 \caption{(a) The DFT-calculated low-energy bands for the twisted BLG/hBN structure with a twist angle of $10.89^{\circ}$. The grey-shaded region indicates the energy window, where we calculated the ILDOS. (b) The density in real-space is of $p_z$-character and localized on the non-dimer atoms of the lower graphene layer. (c) Top view of the structure, where the grey-shaded circles identify different local high-symmetry stackings. (d) Colormap of the valence band energy around the K point. The white dashed lines represent the edge of the Brillouin Zone with the K point at the center. (e) Same as (d), but for the conduction band. (f-h) The atom resolved DOS of the three high-symmetry stackings as identified in the twisted structure in (c). 
 }\label{Fig:ldos_bands}
\end{figure*}

In order to emphasize the local band gap variation, we considered the smallest twisted BLG/hBN structure with a twist angle of $10.89^{\circ}$ as an exemplary case, for which we have calculated the integrated local density of states (ILDOS) for the valence band edge states, see Fig.~\ref{Fig:ldos_bands}(a,b). 
The valence band edge is formed by the non-dimer atoms of the lower graphene layer in direct contact with the hBN, which is nicely reflected in the real-space picture of the ILDOS.
At first glance, one may think that all the relevant non-dimer C atoms contribute equally. However, within the twisted structure, see Fig.~\ref{Fig:ldos_bands}(c), one can identify local high-symmetry stacking configurations, which provide very different band gaps and DOS contributions, see Fig.~\ref{Fig:ldos_bands}(f-h). In other words, by having knowledge about local high-symmetry stackings, one can map the local band gaps to the twisted heterostructures. 
Looking at the band edge energies in the vicinity of the K point, see Fig.~\ref{Fig:ldos_bands}(d,e), also the trigonal warping can be seen~\cite{McCann2013:RPP,Konschuh2012:PRB}. Moreover, the trigonal warping points are no longer directed along high-symmetry lines due to the twist angle. This is also the reason, why we extracted the band gaps for the twisted structures from the band edge extrema around the K point, see Fig.~\ref{Fig:ldos_bands}(d,e), and not from the dispersion along the high-symmetry lines, see Fig.~\ref{Fig:ldos_bands}(a).

\section{Summary}
\label{sec:summ}
In summary, we have calculated the electronic structures of hBN encapsulated BLG from first principles. 
By employing a model Hamiltonian, we were able to reproduce the relevant low-energy bands of BLG using reasonable fit parameters. The main effect of hBN on the BLG dispersion is on the orbital level, introducing band gaps up to tens of meV, depending on the stacking and encapsulation scenario. 
Additionally, while SOC parameters stay in the range of few to tens of $\mu$eV, they are markedly proximity modified by the surrounding hBN layer(s). Further tunability, mainly of the orbital gap and the layer-resolved Rashba SOCs, is provided by a transverse electric field. 
From the investigated twisted BLG/hBN heterostructures, we find that rather large band gap variations arise from the different local stacking configurations.

The modulation of the band gap due to different stacking configurations occurs on a spatial scale on the order of nanometers.  Such precise spatial resolution can be achieved, for example, by  scanning tunneling microscopy (STM), which have been used to characterize BLG on hBN~\cite{Yankowitz_2014Sep, Holdman_2019Oct}. 
In contrast, STM measurements require the BLG to be exposed without encapsulation in hBN. This exposure can lead to contamination and unintentional doping of the top BLG layer, obscuring the impact of diverse stacking orders on the band structure due to variations in disorder potential. Furthermore, the absence of metallic or graphite bottom gates can result in further interference from the substrate. 
Another potential experiment with sufficient spatial resolution could be scanning gate microscopy, which furthermore allows encapsulation of the BLG in hBN. Yet, previous measurements on dual-gated BLG were influenced by localized states~\cite{Gold_2020Dec} that might also mask the variations in the band gap across the sample. 
While transport spectroscopy~\cite{Overweg_2018, Icking2022:AEM} or optical experiments~\cite{Mak_2009, Kuzmenko_2009Mar, Zhang_2009Jun} can determine the band gap in BLG, they yield an averaged gap across the sample due to factors like the laser beam spot size. Still, they could be utilized to study the effect of different twist angles between the hBN and BLG on the effective band gap.

The sensitivity of stacking configuration related spatial band gap variations to external parameters such as strain, interface disorder, or unintentional doping, necessitates samples of exceptionally high quality, particularly relying on state-of-the-art graphite-gated BLG devices.
The presented results should be particularly helpful in understanding and analyzing spectra of BLG quantum point contacts~\cite{Banszerus_2020,Lee_2020} and quantum dots~\cite{Knothe2020:PRB,Moeller2021:PRL,Moeller2023:arxiv,Eich_2018} for building moir\'{e} potentials to realistically model orbital and spin-orbit effects in extended geometries.


\acknowledgments
This work was funded by the European Union Horizon 2020 Research and Innovation Program under contract number 881603 (Graphene Flagship), Deutsche Forschungsgemeinschaft (DFG, German Research Foundation) SFB 1277 (Project No. 314695032), and DFG SPP 2244 (Project No. 443416183).

\bibliography{paper}

\end{document}